\documentclass[prd,aps,floats,epsfig,twocolumn]{revtex4}
\usepackage{graphicx}
\newcommand{\beq}{\begin{equation}}
\newcommand{\eeq}{\end{equation}}
\newcommand{\bea}{\begin{eqnarray}}
\newcommand{\eea}{\end{eqnarray}}

\newcommand{\ApJ}{{\it Astrophys. J.\,}}

\newcommand{\PR}{{\it Phys. Rev.\,}}

\begin{document}

\title{Cosmological constraints in the presence of ionizing and resonance radiation at recombination}
\author{Rachel Bean$^\sharp$, Alessandro Melchiorri$^{*}$ and Joseph Silk
$^\flat$}
\affiliation{$^\sharp$ Dept. of Astronomy, Space Sciences Building, Cornell
  University, Ithaca, NY, USA.\\
$^*$ Dipartimento di Fisica e sezione INFN, Universita' di Roma ``La Sapienza'', Ple Aldo Moro 2, 00185, Rome, Italy.\\
$^\flat$ Astrophysics, Denys Wilkinson Building, University of Oxford,
  Keble Road, OX3RH, Oxford, UK.}

\begin{abstract}

With the recent measurement of full sky cosmic microwave background polarization from WMAP, key cosmological degeneracies have been broken, allowing tighter constraints to be placed on cosmological parameters inferred assuming a standard recombination scenario. Here we consider the effect on cosmological constraints if additional ionizing and resonance radiation sources are present at recombination. We find that the new CMB data significantly improve the
constraints on the additional radiation sources, with $\log_{10}[\epsilon_{\alpha}] < -0.5$ and  $\log_{10}[\epsilon_{i}] <-2.4$
at $95 \%$ c.l. for resonance and ionizing sources respectively. 
Including the generalized recombination scenario, however, we find that the constraints on the scalar spectral index $n_s$ are weakened to $n_s=0.98\pm0.03$, with the $n_s=1$ case now well 
inside the $95\%$ c.l.. The relaxation of constraints on tensor modes,
scale invariance, dark energy and neutrino masses are also discussed.

\end{abstract}
\maketitle

\section{Introduction}

The recent measurements of the Cosmic Microwave Background (CMB) flux provided 
by the three year Wilkinson Microwave Anisotropy Probe (WMAP) 
mission (see \cite{wmap3cosm,wmap3pol,wmap3temp,wmap3beam}
have confirmed several of the results already presented in
the earlier data release, but also pointed towards new conclusions.
The better treatment of systematics in large scale 
polarization data, in particular, has now provided a lower
value for the optical depth parameter $\tau$. This, together with 
an improved signal in the temperature data at higher multipoles,
has resulted in a lower value of the 
spectral index parameter $n_s = 0.959\pm0.016$.
A determination of this parameter can play a crucial role in
the study of inflation. Soon after the WMAP data release, several 
papers have indeed investigated the possibility of discriminating between
single-field inflationary models by making use of this new, high
quality, dataset \cite{alabidi,Peiris:2006ug,Parkinson:2006ku,
Pahud:2006kv,Lewis:2006ma,Seljak:2006bg,Magueijo:2006we,Easther:2006tv,kinney06}.
One of the main conclusions of these papers is that some inflationary
models, such as quartic chaotic models of the
form $V(\phi)\sim \lambda \phi^4$, may be considered ruled out by the
current data while others, such as chaotic inflation with a quadratic
potential $V(\phi) \sim m^2 \phi^2$ are consistent with all data sets.
 
While the WMAP result is of great importance for inflationary model
building, one should be careful in taking any conclusion as definitive
since the constraints on $n_s$ are obtained in an indirect way and
are, therefore, model dependent. Similar considerations applies to other cosmological constraints,
such those on the dark energy equation of state and neutrino
masses. Combining CMB anisotropies with 
galaxy clustering and supernovae type Ia data,
the dark energy equation of state parameter 
(dark energy pressure over density) has been constrained to $w=-1.08 \pm0.12$ 
at $95 \%$ c.l. (see \cite{wmap3cosm}). 
Using the same dataset, but under the assumption of a cosmological constant,
it is possible to constrain the neutrino masses
to $\sum m_i < 0.66 eV$ at $95 \%$ c.l. where $i=1,..,3$ and
indicates the neutrino flavor. Again, while those constraints play a 
very important role in our understanding of the dark energy component and 
neutrino physics, they are obtained in an indirect way and under
several assumptions.

The importance of the model dependency 
of the cosmological constraints has been recently discussed by several authors.
The impact of isocurvature modes on the determination
of the neutrino mass \cite{pedro}, dark energy properties
\cite{trotta}, scalar spectral index 
\cite{Bean:2006qz} and baryon density \cite{Keskitalo:2006qv} is just one example.

Here we investigate possible deviations in the mechanism on which CMB 
anisotropies are highly dependent: the process of recombination. 

In a previous paper \cite{bms}, we analyzed modified 
recombination processes in light of the WMAP first year data.
Here we assess the improvements given by more recent data, in
particular the inclusion of CMB polarization spectra, and also
extend the analysis to a larger set of parameters.
We will indeed not only provide new and more stringent constraints
on modified recombination but we  also consider
its impact on inflationary, dark energy and neutrino parameters, 

The recombination process can be modified in several ways. For
example, one could use a model-independent, phenomenological approach 
such as in \cite{Hannestad01} where models are specified by the
position and width of the recombination surface in redshift space. 
Here we instead focus on theoretically motivated mechanisms based on 
extra sources of ionizing and resonance radiation at recombination 
(see e.g. \cite{Seager00}).
While the method we adopt will be general enough to cover
most of the models of this kind, as discussed in the next section, we remind the
reader that there exist other ways in which to modify 
recombination, for instance, by having a time-varying 
fine-structure constant (\cite{alpha}).

Following the seminal papers \cite{Peebles68,Zeldovich68} detailing the 
recombination process, further refinements to the standard scenario 
were developed \cite{Seager99}, allowing predictions at the accuracy level 
found in data from the WMAP satellite and the  future Planck satellite 
\cite{Hu95,Seljak03}. With this level of accuracy, it becomes conceivable 
that deviations from standard recombination maybe be detectable 
\cite{Seager00,Naselsky02,Dorosh02}, although further refinements could be required to get the Thomson visibility function below percent level accuracy \cite{Chluba:2005uz,Chluba:2006bc,Wong:2006iv}.

The paper proceeds as follows: in section \ref{sec2} we describe a model  
which can produce deviations from the standard recombination scenario. 
In \ref{sec3}, we describe how these deviations might affect the 
CMB temperature and polarization power spectra and conduct a likelihood 
analysis using the recent CMB data from WMAP and other cosmological
observables. 
In particular, we will study the impact that a modified recombination 
scheme can have on several cosmological and astrophysical parameters.
In \ref{sec4} we draw together the implications of the analysis.

\section{A modified ionization history}\label{sec2}

The evolution of the ionization fraction, $x_{e}$, of atoms, number density $n$, can be modeled in a simplified manner for the recombination of hydrogen, \cite{Peebles68,Zeldovich68}, 

\bea 
-{dx_{e}\over dt}\left.\right|_{std}=C\left[a_c n x_{e}^{2}-b_c 
(1-x_{e})\exp{\left(-{\Delta B\over k_{B}T}\right)}\right]
\label{eq1}
\eea
where  $a_{c}$ and $b_{c}$ are 
the effective recombination and photo-ionization rates for principle 
quantum numbers $\ge 2$, $\Delta B$ is the difference in binding energy between the $1^{st}$ and $2^{nd}$ energy levels and
\bea C={1+K\Lambda_{1s2s}n_{1s}\over1+K(\Lambda_{1s2s}+b_{c})n_{1s}}
\label{eqC},  \ \ \ \ K={\lambda_\alpha^{3} \over8\pi H(z)}
\eea
where  $\lambda_{\alpha}$ is the wavelength of the single Ly-$\alpha$ transition from the $2p$ level, $\Lambda_{1s2s}$ is the decay rate of the metastable $2s$ level, $n_{1s}=n(1-x_{e})$ is the number of neutral ground state $H$ atoms, and $H(z)$ is the Hubble expansion factor at a redshift $z$.

We include the possibility of extra photons at key wavelengths that would modify this recombination picture, namely, resonance (Ly-$\alpha$) photons with number density, $n_{\alpha}$,which promote electrons to the $2p$ level, and ionizing photons, $n_{i}$,\cite{Seager00,Naselsky02,Dorosh02,bms}
\bea
{dn_{\alpha}\over dt}&=&\varepsilon_{\alpha}(z)H(z)n, \ \ \ 
{dn_{i}\over dt}=\varepsilon_{i}(z)H(z)n.
\label{eq0}
\eea
which leads to a modified evolution of the ionization fraction
\bea -{dx_{e}\over dt}&=&-{dx_{e}\over dt}\left.
\right|_{std}-C\varepsilon_{i} H-(1-C)\varepsilon_{\alpha}H . 
\ \ \ \ \ \ \ \label{eq2}
\eea

Extra photon sources can be generated by a variety of mechanisms. A widely considered process is the decay or annihilation of massive particles \cite{Sarkar:1983,Scott:1991,Ellis:1992,Adams:1998,Seager00,Doroshkevich:2002ff,Naselsky:2003zj,Zhang:2006fr}.  The decay channel depends on the nature of the particles, and could, for example, include charged and neutral leptons, quarks or gauge bosons. These particles may then decay further, leading to a shower /cascade that could, amongst other products, generate a bath of lower energy photons that could interact with the primordial gas and cosmic microwave background.  Interestingly these models, as well as injecting energy at recombination, $z\sim 1000$, boost the ionization fraction after recombination and can distort the ionization history of the universe at even later times, during galaxy formation and reionization $z\sim 5-10$ \cite{Pierpaoli:2003rz,Chen:2003gz,Padmanabhan:2005es,Mapelli:2006ej,Lewis:2006ma}. Other mechanisms include evaporation of black holes \cite{Naselsky87,Naselsky02} or inhomogenities in baryonic matter \cite{Naselsky02}.

We employ the widely used \texttt{RECFAST} code \cite{Seager99}, in the \texttt{cosmomc} package \cite{Lewis:2002ah} modifying the code as in (\ref{eq2}) to include two extra constant parameters, $\epsilon_{\alpha}$ and $\epsilon_{i}$. In addition to the ionizing sources, we assume a single, swift reionization epoch at a redshift $z_{re}$. 

\begin{figure*}[t]
\begin{center}
\includegraphics[width=6.5in]{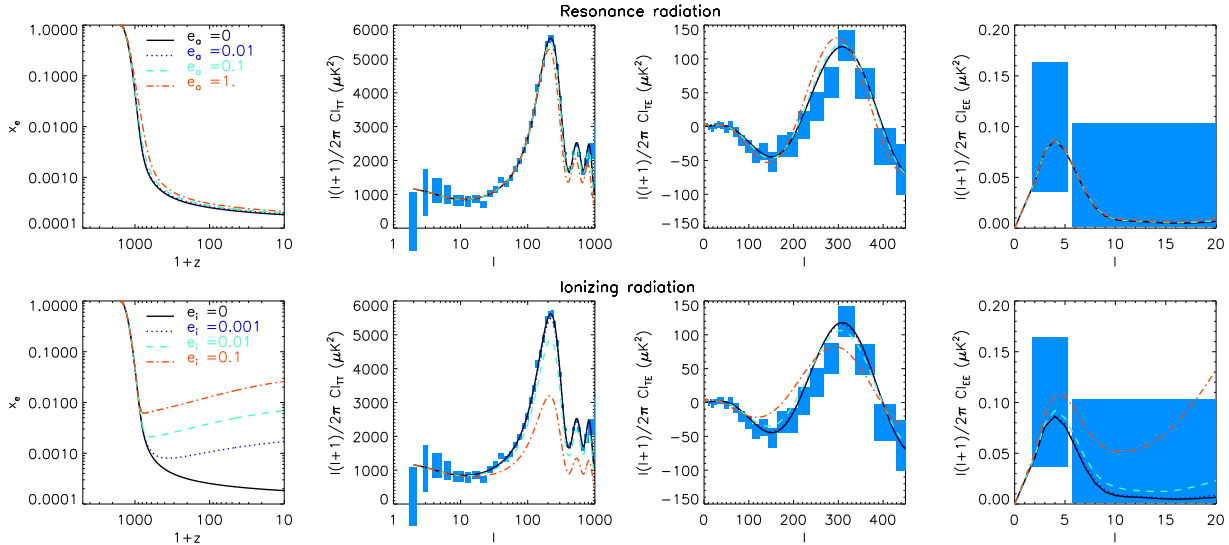}
\caption{(From left to right) The comparison of ionization fraction evolution, and TT (left) ,TE (center) and EE (right) CMB spectra comparing a best fit $\Lambda$CDM fiducial model to models with the same cosmological parameters but with additional resonance (top) and ionizing (bottom) radiation. WMAP binned data are shown as blue shaded regions. }
\label{cls}
\end{center}
\end{figure*}

In Figure \ref{cls} we show the effect of additional resonance and ionizing radiation on the CMB TT, TE and EE spectra, in comparison to a fiducial best fit model to the WMAP 3-year data. From identical initial power spectra, the inclusion of additional resonance photons slighly boosts the ionization fraction at and after recombination, suppressing TT power at small scales, while the large scale EE spectra is largely unaffected. Ionizing photons significantly boost the ionization fraction post recombination and therefore as well as significantly suppressing TT power on small scales, they also can generate a boost in the large scale EE signal akin to an early partial reionization. 

\section{Likelihood analysis}
\label{sec3}

The method we adopt is based on the publicly available Markov Chain Monte Carlo
package \texttt{cosmomc} \cite{Lewis:2002ah}. We sample the following
dimensional set of cosmological parameters, adopting flat priors on them:
the physical baryon and Cold Dark Matter (CDM) densities, $\omega_b=\Omega_bh^2$ and
$\omega_c=\Omega_ch^2$, the ratio of the sound horizon to the angular diameter
distance at decoupling, $\theta_s$, the scalar spectral index, $n_{s}$,
and the optical depth to reionization, $\tau$.  As described in the
previous section, we modify recombination by considering
variations in the $\varepsilon_{\alpha}$ and $\varepsilon_{i}$
parameters. Furthermore, we consider purely adiabatic
initial conditions and we impose flatness. We also consider the
possibility of having a tensor (gravity waves) 
component with amplitude $r$ respect to scalar, a running of the
spectral index $dn_{s}/dlnk$ at $k=0.002 h^{-1}Mpc$ and
a non-zero, degenerate, neutrino mass of energy density:

\bea
\Omega_{\nu}h^2=\frac{\Sigma m_{\nu}}{92.5 eV}
\eea

\noindent Finally, we will also investigate the possibility
of a dark energy equation of state, $w$, different from $-1$ but
constant with redshift.
The MCMC convergence diagnostics is done on $7$ chains though the
Gelman and Rubin ``variance of chain mean''$/$``mean of chain variances'' $R$
statistic for each parameter. Our $1D$ and $2D$ constraints are obtained
after marginalization over the remaining ``nuisance'' parameters, again using
the programs included in the \texttt{cosmomc} package. In addition to the WMAP
data, we also consider the constraints on the real-space power spectrum of
galaxies from the Sloan Digital Sky Survey (SDSS) \cite{Tegmark:2003uf}. We restrict the
analysis to a range of scales over which the fluctuations are assumed to be in
the linear regime ($k < 0.2 h^{-1}$\rm Mpc). When combining the matter power
spectrum with CMB data, we marginalize over a bias $b$ considered as an
additional nuisance parameter. Furthermore, we make use of the Hubble Space Telescope (HST) measurement
of the Hubble parameter $H_0 = 100h \text{ km s}^{-1} \text{Mpc}^{-1}$
\cite{hst} by multiplying the likelihood by a Gaussian likelihood function
centered around $h=0.72$ and with a standard deviation $\sigma
=0.08$. When considering dark energy models, we also
include information from luminosity distance measurements of
type Ia Supernovae from the recent analysis of \cite{astier}.
Finally, we include a top-hat prior on the age of the universe: 
$10 < t_0 < 20$ Gyrs.

\section{Results}
\label{sec4}

\begin{figure}[t]
\begin{center}
\includegraphics[width=3.2in]{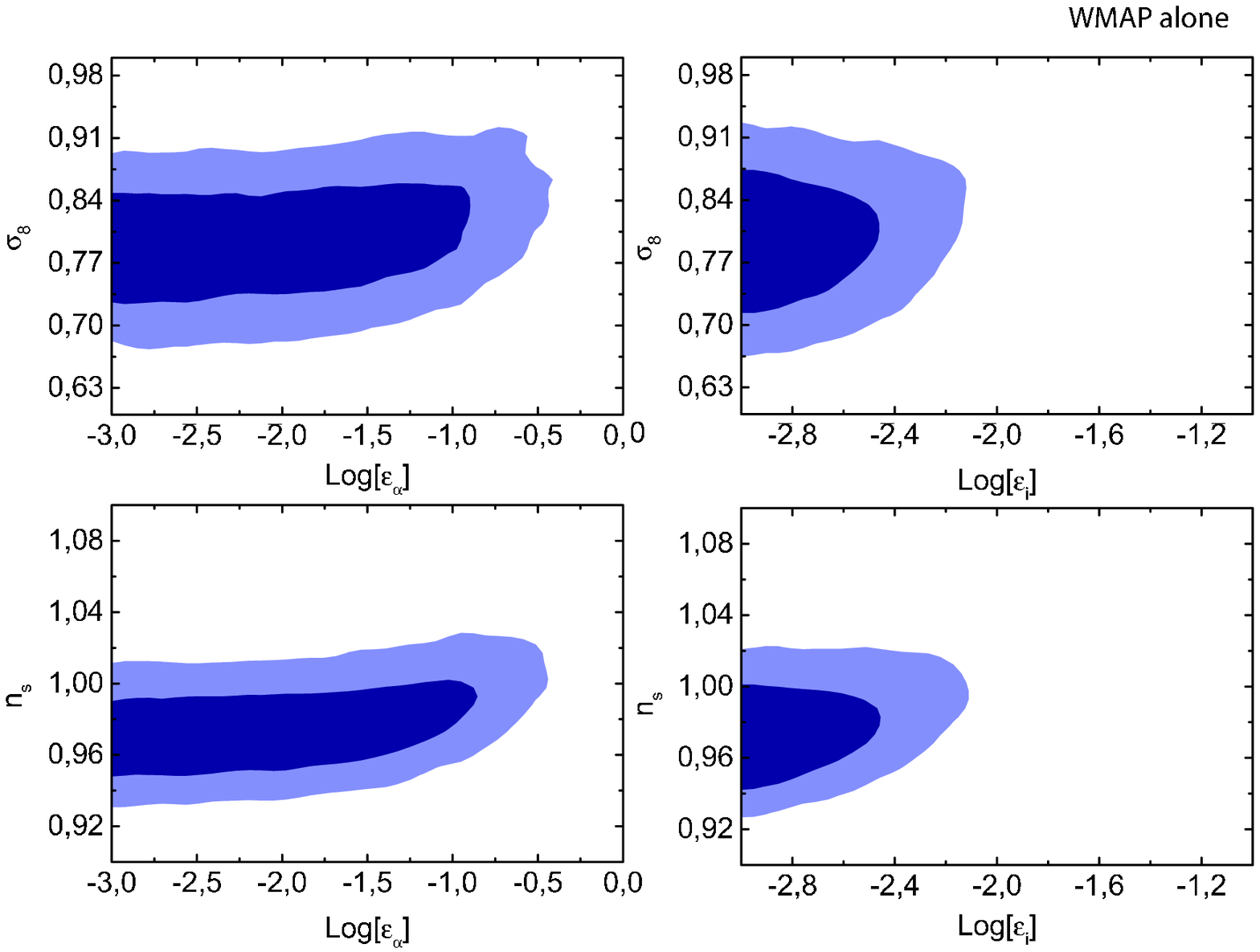}
\includegraphics[width=3.2in]{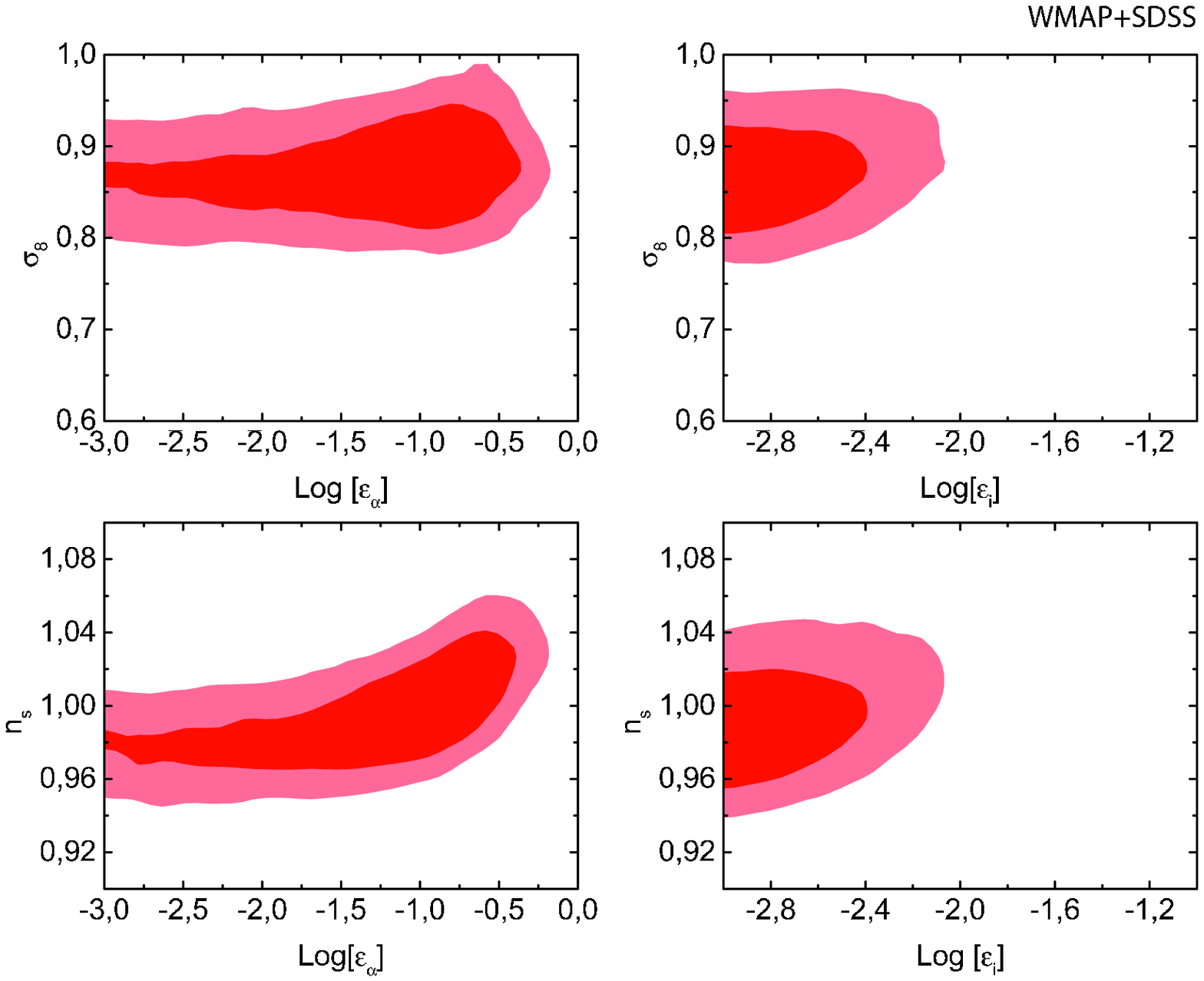}
\caption{The $68 \%$ and $95 \%$ likelihood contours in the 
$n_s$ and $\sigma_8$ vs $\varepsilon_{\alpha}$ plane (left)
  and $n_s$ and $\sigma_8$ vs $\varepsilon_{i}$ (right).
The analysis considers (top/blue) the $3$-years WMAP data and 
a HST prior on the Hubble parameter, $h$, alone and (bottom/red) also including SDSS galaxy matter power spectrum data.}
\label{twod1}
\end{center}
\end{figure}

Our main results are plotted in Figure \ref{twod1}
where we show the $68 \%$ and 
$95 \%$ c.l. on the $n_s-\log_{10}(\epsilon_\alpha)$,
$\sigma_8-\log_{10}(\epsilon_\alpha)$, $n_s-\log_{10}(\epsilon_i)$ and 
$\sigma_8-\log_{10}(\epsilon_i)$ plane.
In the top portion of Figure \ref{twod1}, we consider only the WMAP data (plus a prior on the
Hubble parameter), while in the lower portion, we add SDSS.
Let us first consider the case of WMAP alone.
As we see, using this dataset alone, we can put interesting
new bounds on the recombination parameters.
Marginalizing over the remaining, ``nuisance'', parameters
we indeed obtain  $\log_{10}(\epsilon_\alpha)<-0.81$ 
and  $\log_{10}(\epsilon_i)<-2.31$ at $95 \%$ c.l..

As suggested by Figure \ref{cls}, we find ionizing photons are better constrained with current data since the ionization fraction is significantly boosted at and beyond the onset of recombination. This results in a suppression of  TT power and boosting of EE power even on large scales, well constrained by WMAP data. Resonance photons have a more subtle effect only slightly increasing the ionization fraction after the onset of recombination. This leads to a suppression of small scale TT power but little effect on large scale EE. The constraints on both types of radiation should be noticeably improved therefore by future experiments, such as the planned PLANCK satellite, which better characterize small scale TT and EE power \cite{Dorosh02}.

Moreover, there is a clear degeneracy between $\log_{10}(\epsilon_\alpha)$ 
and  the spectral index $n_s$. Indeed, a modification of the
recombination scheme allows us to suppress the
amplitude of the peaks in the CMB power spectrum 
in a way similar to a later recombination but without 
altering the large-scale polarization signal.
This changes in a drastic way the constraints on the
scalar spectral index and $\sigma_8$. Marginalizing over 
the recombination parameters, we get 
$n_s=0.978_{-0.029}^{+0.032}$ and 
$\sigma_8=0.80_{-0.09}^{+0.08}$ at $95 \%$ c.l.. 
Those results should be compared with the constraints
$n_s=0.959_{-0.027}^{+0.026}$ and 
$\sigma_8=0.78_{-0.07}^{+0.08}$, again at $95 \%$ c.l., 
obtained using the same
dataset and priors but with standard recombination.

Including SDSS data, as is shown in the lower panel in Figure \ref{twod1}, does not significantly improve our constraints on 
 $\varepsilon_{\alpha}$ and  $\varepsilon_{i}$.
The SDSS power spectrum indeed prefers
a higher value of the $\sigma_8$ parameter
than WMAP. While the tension is not strong
enough to provide any evidence for modified 
recombination, the constraints are lowered
to  $\log_{10}(\epsilon_\alpha)<-0.51$ 
for  $\varepsilon_{\alpha}$ and almost stable
to $\log_{10}(\epsilon_i)<-2.24$, 
for  $\varepsilon_{i}$ at $95 \%$ c.l..
The constraints on $n_s$ and $\sigma_8$ are
also affected. Including SDSS we find $n_s=0.994_{-0.035}^{+0.040}$ and 
$\sigma_8=0.87_{-0.06}^{+0.07}$ at $95 \%$ c.l..

\begin{figure}[t]
\begin{center}
\includegraphics[width=2.9in]{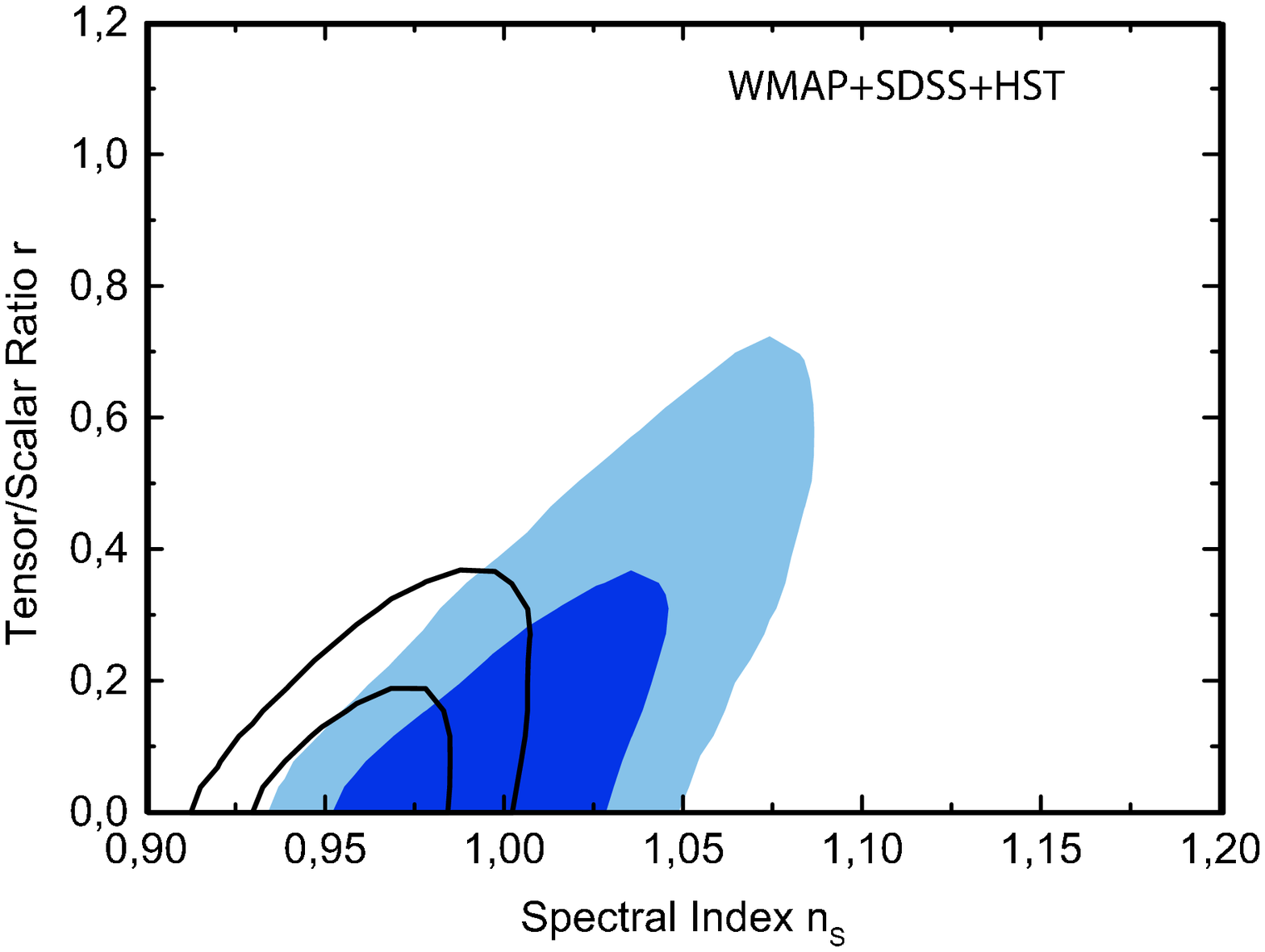}
\includegraphics[width=3.0in]{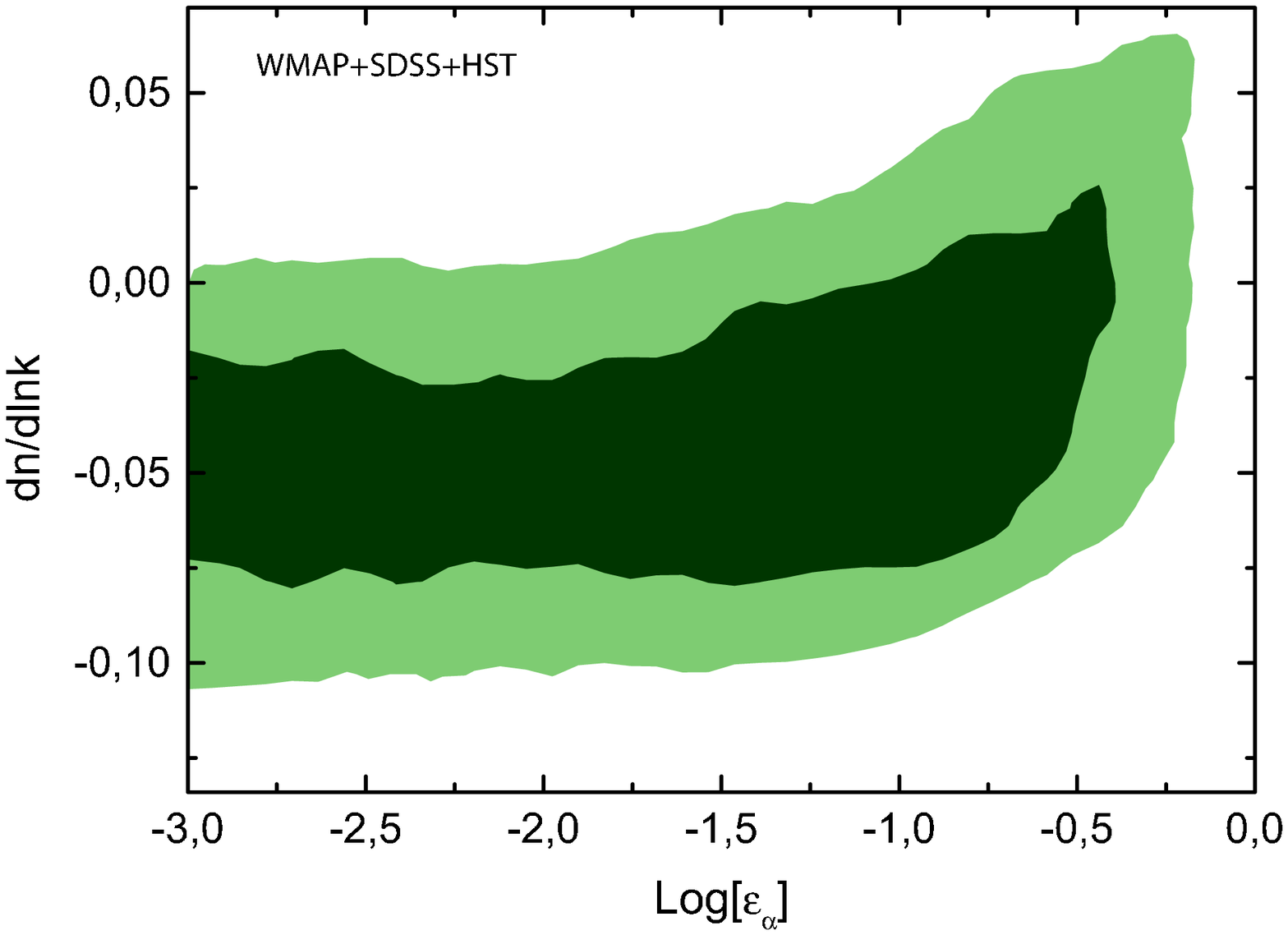}
\caption{The effect of a modified recombination scheme
on inflationary parameters in a WMAP+SDSS analysis.
In the top panel we plot the constraints in the $n_s-r$ plane. The filled contours ($68 \%$ and $95 \%$) 
are obtained after marginalization over extended recombination
parameters  while the empty contours assume standard recombination.
In the bottom panel, we show the $68 \%$ and $95 \%$ likelihood contours in the 
$dn_{s}/dlnk$ vs $\varepsilon_{\alpha}$ plane. Non-standard 
recombination shifts the $\sim 1 \sigma$ evidence
for running in the standard case to a null result.
Future evidence for running may be interpreted
as evidence for a non-standard recombination scheme.}
\label{twod2}
\end{center}
\end{figure}

\begin{figure}[t]
\begin{center}
\includegraphics[width=3.0in]{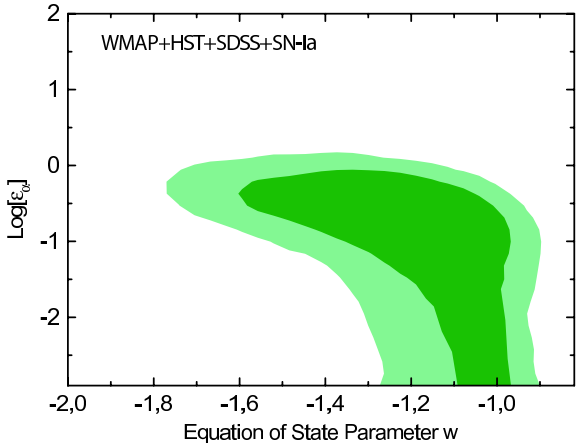}
\caption{The impact of a modified recombination scheme
on constraining a constant dark energy equation of state, $w$.
We show the $68 \%$ and $95 \%$ likelihood contours in the 
$w$ vs $\varepsilon_{\alpha}$ plane from WMAP+SDSS+HST+SN-1a (see
text). Non-standard 
recombination relaxes the constraints towards more negative values for $w$}
\label{twod3}
\end{center}
\end{figure}

\begin{figure}[t]
\begin{center}
\includegraphics[width=3.0in]{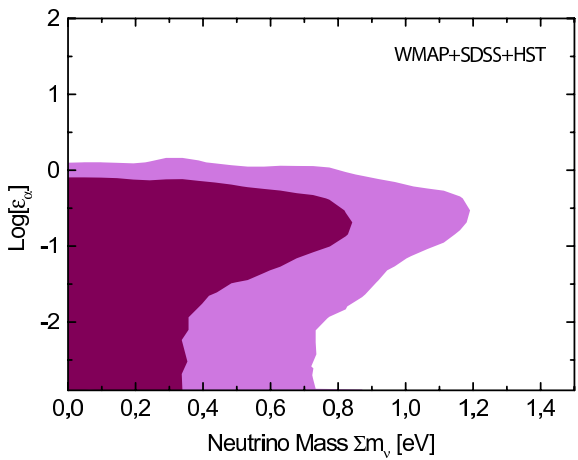}
\caption{The effect of a modified recombination scheme
on constraining neutrino masses.
We show the $68 \%$ and $95 \%$ likelihood contours in the 
$\Sigma m_{\nu}$ vs $\varepsilon_{\alpha}$ plane from WMAP+SDSS+HST (see
text). Non-standard 
recombination relax the constraints towards larger masses}
\label{twod4}
\end{center}
\end{figure}
Interestingly, we find that the constraints on other key
parameters ($\tau$ or $\Omega_b$) are 
robust to the modifications in the recombination scenario.
It is interesting to extend the analysis to
other inflationary parameters such as the amplitude
of a tensor component $r$ or a running of the spectra
index $dn_{s}/dlnk$. In Figure \ref{twod2} (top panel) 
we plot the $68 \%$ and $95 \%$ likelihood contours 
in the $n_s-r$ plane in the standard and in the
generalized recombination case. As one can see, 
relaxing our knowledge about recombination 
strongly affects the final constraints: 
the scalar spectral index can be more consistent 
with $n_s > 1$ and the upper limit on the tensor
component can be a factor $2$ larger than in the 
standard case.
As one can see from the bottom panel of Figure \ref{twod2}
a degeneracy between $\epsilon_{\alpha}$
and the running $dn_{s}/dlnk$ is also present.
Standard analyses prefer a negative running of
the spectral index with significance slightly above
$1 \sigma$ (see \cite{wmap3cosm}). This can be compensated for by 
a non-standard recombination with
 $\epsilon_{\alpha} > 0.1$.

In Figure \ref{twod3}, we report on the impact of non-standard recombination on
the equation of state parameter, $w$.
We find an important degeneracy only with $\epsilon_{\alpha}$;
allowing $\epsilon_{\alpha}$ to vary enlarges the constraints
on $w$ towards more negative values.
A future, combined, indication for $w<-1$ could,
therefore, provide a hint of a non-standard recombination
process and one should be careful in interpreting it 
as evidence for a phantom-like dark energy component.
In a more generalized recombination scenario, we find
the constraints on $w$ are relaxed to
$w=-1.24_{-0.44}^{+0.28}$ at $95 \%$ c.l..

Finally, in Figure \ref{twod4}, we report the constraints on
neutrino masses. As one can see, non-standard recombination
 also relaxes constraints on this parameter.
We find that values as large as 
$\Sigma m_{\nu} \sim 1.2 eV$ are consistent with the
data, relaxing by $\sim 50 \%$ the standard constraint 
 $\Sigma m_{\nu}< 0.72 eV$ (see e.g. \cite{wmap3cosm,fogli06}).

\section{Conclusions}
\label{sec5}

In this paper, we update the upper bounds that can be placed on the contribution 
of extra Ly-$\alpha$ and ionizing photon-producing sources in light of
the new WMAP data.  
We find that, adopting a simple parametrization using constant
effective values for $\varepsilon_{\alpha}$ and $\varepsilon_{i}$,
the WMAP data constraints $\log_{10}[\varepsilon_{\alpha}] <-0.5$ and 
$\log_{10}[\varepsilon_{i}]<{-2.4}$ at the $95\%$ level. Physically motivated models for non-standard recombination which generate ionizing and resonance radiation, like those 
based on primordial black hole or super-heavy dark matter decay, remain
feasible. 

We find that a modified recombination scheme may affect the
current WMAP constraints on inflationary parameters like the spectral
index $n_{s}$ and its running. In particular, if recombination is
changed, Harrison-Zel'dovich spectra with $n_s=1$, larger tensor modes and
positive running are in agreement with observations. Moreover, constraints on particle physics parameters like the neutrino mass are also relaxed when non-standard recombination is considered. 

Future observations in both temperature and polarization, 
such as those expected from the Planck satellite \cite{Dorosh02}, 
will provide more precise small scale TT and EE measurements needed to more stringently test these models and, crucially, will reduce the dependency of other cosmological parameters on them.

\medskip 
\textbf{Acknowledgments} 

RB's work is supported by NSF grants AST-0607018 and PHY-0555216 and uses National Supercomputing (NCSA) resources under grant TG-AST060029T.


\begin{thebibliography}{99}


\bibitem{wmap3cosm}
D. N. Spergel {\it et al.},
arXiv:astro-ph/0603449.

\bibitem{wmap3pol}
L. Page {\it et al.},
arXiv:astro-ph/0603450.

\bibitem{wmap3temp} G. Hinshaw {\it et al.},
arXiv:astro-ph/0603451.

\bibitem{wmap3beam}
N. Jarosik {\it et al.},
arXiv:astro-ph/0603452.


\bibitem{alabidi}
L. Alabidi and D. H. Lyth,
arXiv:astro-ph/0603539.

\bibitem{Peiris:2006ug}
H. Peiris and R. Easther,
arXiv:astro-ph/0603587.

\bibitem{Parkinson:2006ku}
D. Parkinson, P. Mukherjee and A. R. Liddle,
arXiv:astro-ph/0605003.

\bibitem{Pahud:2006kv}
C. Pahud, A. R. Liddle, P. Mukherjee and D. Parkinson, arXiv:astro-ph/0605004.

\bibitem{Lewis:2006ma}
A. Lewis,
arXiv:astro-ph/0603753.

\bibitem{Seljak:2006bg}
U. Seljak, A. Slosar and P. McDonald,
arXiv:astro-ph/0604335.

\bibitem{Magueijo:2006we}
J. Magueijo and R. D. Sorkin,
arXiv:astro-ph/0604410.

\bibitem{Easther:2006tv}
R. Easther and H. Peiris,
arXiv:astro-ph/0604214.

\bibitem{kinney06}
W.~H.~Kinney, E.~W.~Kolb, A.~Melchiorri and A.~Riotto,
  arXiv:astro-ph/0605338; PRD in press.

\bibitem{pedro}
  C.~Zunckel and P.~G.~Ferreira,
  arXiv:astro-ph/0610597.

\bibitem{trotta}
  R.~Trotta, A.~Riazuelo and R.~Durrer,
  Phys.\ Rev.\ D {\bf 67} (2003) 063520
  [arXiv:astro-ph/0211600].

\bibitem{Bean:2006qz}
  R.~Bean, J.~Dunkley and E.~Pierpaoli,
  Phys.\ Rev.\ D {\bf 74}, 063503 (2006)
  [arXiv:astro-ph/0606685].

\bibitem{Keskitalo:2006qv}
  R.~Keskitalo, H.~Kurki-Suonio, V.~Muhonen and J.~Valiviita,
  arXiv:astro-ph/0611917.

\bibitem{bms}
  R.~Bean, A.~Melchiorri and J.~Silk,
  Phys.\ Rev.\ D {\bf 68} (2003) 083501
  [arXiv:astro-ph/0306357].


\bibitem{Hannestad01} S. Hannestad \& R. J.  Scherrer, \PR D {\bf 63}, 083001 (2001)
\bibitem{Seager00} P.J.E. Peebles, S. Seager, W. Hu, \ApJ {\bf 539} L1 (2000), astro-ph/0004389. 
\bibitem{alpha} S. Hannestad, \PR D {\bf 60}, 023515 (1999); 
M. Kaplinghat, R. J. Scherrer \& M S Turner, \PR D {\bf 60},023516 (1999); 
P. P. Avelino et al.,PRD {\bf 62}, 123508 (2000); 
R. Battye, R. Crittenden \& J. Weller, \PR D {\bf 63}, 043505 (2001); 
P.~P.~Avelino {\it et al.},
\PR D {\bf 64} (2001) 103505
[arXiv:astro-ph/0102144]
Landau, Harari \& Zaldarriaga, \PR D {\bf 63}, 083505 (2001).
  C.~J.~A.~Martins, A.~Melchiorri, G.~Rocha, R.~Trotta, P.~P.~Avelino and P.~Viana,
  Phys.\ Lett.\ B {\bf 585}, 29 (2004)
  [arXiv:astro-ph/0302295].


\bibitem{Peebles68} P.J.E. Peebles, \ApJ {\bf 153} 1 (1968).
\bibitem{Zeldovich68} Ya. B. Zel'dovich, V.G. Kurt, R.A. Sunyaev, {\it Zh. Eksp. Teoret. Fiz} {\bf 55} 278(1968), English translation, {\it Sov. Phys. JETP.} {\bf 28} 146 (1969).
\bibitem{Seager99} S. Seager, D.D. Sasselov, \& D. Scott, \ApJ {\bf 523} 1 (1999), astro-ph/9909275.
\bibitem{Hu95} W. Hu, D. Scott, N. Sugiyama, \& M. White, \PR D {\bf 52} 5498 (1998).
\bibitem{Seljak03}  U. Seljak, N. Sugiyama, M. White, M. Zaldarriagaastro-ph/0306052

\bibitem{Naselsky02} P.D. Naselsky, I.D. Novikov {\it MNRAS} {\bf 334} 137 (2002), astro-ph/0112247
\bibitem{Dorosh02} A.G. Doroshkevich, I.P. Naselsky, P.D. Naselsky, I.D. Novikov, \ApJ {\bf 586} 709 (2002), astro-ph/0208114.



\bibitem{Chluba:2005uz}
  J.~Chluba and R.~A.~Sunyaev,
  Astron.\ Astrophys.\  {\bf 446}, 39 (2006)
  [arXiv:astro-ph/0508144].

\bibitem{Chluba:2006bc}
  J.~Chluba, J.~A.~Rubino-Martin and R.~A.~Sunyaev,
  arXiv:astro-ph/0608242.

\bibitem{Wong:2006iv}
  W.~Y.~Wong and D.~Scott,
  arXiv:astro-ph/0610691.


\bibitem{Sarkar:1983} S.~Sarkar and A.~Cooper, Phys. Lett. B., {\bf 148}, 347 (1983)
\bibitem{Scott:1991} D. Scott ,M.~J.~ Rees  \& D.~W.~ Sciama , A \& A, {\bf 250}, 295, (1991)
\bibitem{Ellis:1992} J. Ellis, G. Gelmini , J. Lopez , D. Nanopoulos, \& S. Sarkar , Nucl.Phys.B. {\bf 373} 399 (1992).
\bibitem{Adams:1998} A. J. Adams J.A., S. Sarkar \& D.W. Sciama , MNRAS, {\bf 301}, 210 (1998)


\bibitem{Doroshkevich:2002ff}
  A.~G.~Doroshkevich and P.~D.~Naselsky,
  Phys.\ Rev.\ D {\bf 65}, 123517 (2002)
  [arXiv:astro-ph/0201212].

\bibitem{Naselsky:2003zj}
  P.~D.~Naselsky and L.~Y.~Chiang,
  Phys.\ Rev.\ D {\bf 69}, 123518 (2004)
  [arXiv:astro-ph/0312168].

\bibitem{Zhang:2006fr}
  L.~Zhang, X.~L.~Chen, Y.~A.~Lei and Z.~G.~Si,
  Phys.\ Rev.\ D {\bf 74}, 103519 (2006)
  [arXiv:astro-ph/0603425].



\bibitem{Pierpaoli:2003rz}
  E.~Pierpaoli,
  Phys.\ Rev.\ Lett.\  {\bf 92}, 031301 (2004)
  [arXiv:astro-ph/0310375].

\bibitem{Chen:2003gz}
  X.~L.~Chen and M.~Kamionkowski,
  Phys.\ Rev.\ D {\bf 70}, 043502 (2004)
  [arXiv:astro-ph/0310473].

\bibitem{Padmanabhan:2005es}
  N.~Padmanabhan and D.~P.~Finkbeiner,
  Phys.\ Rev.\ D {\bf 72}, 023508 (2005)
  [arXiv:astro-ph/0503486].

\bibitem{Mapelli:2006ej}
  M.~Mapelli, A.~Ferrara and E.~Pierpaoli,
  Mon.\ Not.\ Roy.\ Astron.\ Soc.\  {\bf 369}, 1719 (2006)
  [arXiv:astro-ph/0603237].

\bibitem{Lewis:2006ym}
  A.~Lewis, J.~Weller and R.~Battye,
  Mon.\ Not.\ Roy.\ Astron.\ Soc.\  {\bf 373}, 561 (2006)
  [arXiv:astro-ph/0606552].

\bibitem{Naselsky87} P.D. Naselsky \& A.G Polnarev, {\it Sov. Astron. Lett.} {\bf 13} 67 (1987)

\bibitem{Lewis:2002ah}
A. Lewis and S. Bridle,
Phys.\ Rev.\ D {\bf 66}, 103511 (2002) (Available from 
\texttt{http://cosmologist.info}.)

\bibitem{Tegmark:2003uf}
  M.~Tegmark {\it et al.}  [SDSS Collaboration],
  Astrophys.\ J.\  {\bf 606}, 702 (2004)
  [arXiv:astro-ph/0310725].

\bibitem{hst}
  W.L. Freedman {\it et al.}, Astrophys. J. {\bf 553}, 47 (2001).

\bibitem{astier}
  P.~Astier {\it et al.},
  Astron.\ Astrophys.\  {\bf 447} (2006) 31
  [arXiv:astro-ph/0510447].

\bibitem{fogli06}
  G.~L.~Fogli {\it et al.},
  arXiv:hep-ph/0608060.


\end{thebibliography}
\end{document}